\documentstyle[prl,aps]{revtex} 
\begin{document} 
\draft 
\title{Phase transitions in periodically driven macroscopic systems}
\author{Sreedhar B. Dutta} 
\address{Tata Institute of Fundamental
Research, Homi Bhabha Road, Mumbai 400 005, India} 
\date{December 1, 2003}
\maketitle 
\begin{abstract} 
We study the large-time behavior of a class of periodically driven
macroscopic systems. We find, for a certain range of the parameters of
either the system or the driving fields, the time-averaged asymptotic
behavior effectively is that of certain other equilibrium systems.
We then illustrate with a few examples how the conventional knowledge of
the equilibrium systems can be made use in choosing the driving fields to
engineer new phases and to induce new phase transitions.
\end{abstract}
\pacs{PACS numbers: 05.70.Fh, 64.60.Ht, 02.50.Ey}

\section{Introduction}
\label{intro}
With an enduring interest studies have been pursued to understand systems
with many degrees of freedom at, near, and far away from equilibrium.
Systems at equilibrium are studied within the well-established 
Boltzmann-Gibbs framework, while those near and far from equilibrium
lack such a framework to study within and are usually described by
stochastic dynamical equations. The stochastic equations that describe the
dynamics near equilibrium need to ensure that the system relaxes to
equilibrium and hence are less difficult to construct than when the system
is far away from equilibrium. However this condition, that the asymptotic
solution is the equilibrium distribution, does not provide a unique
stochastic equation. In fact, for a given equilibrium system infinitely
many such equations near equilibrium can be provided that would lead to
the same static properties\cite{zinn}. This naturally motivates the study
of various stochastic models. 
 
We know that a closed system, characterized by the Hamiltonian $H$, 
when exposed to the environment for a long time, will equilibriate
and be described by the Boltzmann distribution $\exp (-\beta H)$.  
If this system is also subjected to an external periodic force then what
is its large-time behavior? Due to the absence of an established framework
to find this asymptotic behavior it is essential to study various
periodically driven stochastic systems. It is also essential as some of
these systems show interesting behavior and some of them furnish
useful applications. Stochastic resonance\cite{gammaitoni} and magnetic
hysteresis\cite{chakra} are some of the well known phenomena that are
exhibited by certain periodically driven stochastic systems.

A large number of stochastic processes have been studied for many decades
now\cite{kampen}. Studies in critical dynamics have been actively
pursued\cite{dominicis} after renormalization group techniques were
successfully applied to equilibrium systems\cite{statfield}. Many driven
diffusive models were studied\cite{schmit} and the critical behavior of
some of them were analyzed too\cite{janssen,leung}. Periodically driven
stochastic particle systems were extensively studied in the context
of stochastic resonance\cite{gammaitoni}, while similar studies on fields
were relatively few. 

The issues that we address in this paper are related to a class of 
periodically driven macroscopic systems that, in the absence of
driving, would relax to equilibrium. The relevant degrees are described by
a field and the equilibrium properties are characterized by an energy
functional of the field. These systems may or may not have a critical
point when in equilibrium but could exhibit a critical behavior when the
driving fields are switched on. Some of the questions that we ask: How do
these macroscopic systems respond to periodically driven fields? How do we
describe the phases of the driven system? What is the dependence of the
critical points, if any, on the parameters of the driving fields? Can
driving change the nature of the phase transitions? Can it lead to new
phases? 

The layout of the paper is as follows. In the next section, we define a
class of stochastic models to describe the relevant fields of periodically
driven systems. In Sec.\ref{asymp}, we develop a perturbative scheme to
solve the Fokker-Planck equation that describes the systems of our
interest and explicitly find their asymptotic behavior to first order. In
Sec.\ref{examples}, we illustrate with examples some of the effects of
driving on phases and phase transitions.

\section{Stochastic Models for Periodically Driven Systems}
\label{models}
In this section, we will define a class of periodically driven stochastic  
systems. We will assume the slowly relaxing modes of the system in the
absence of the periodic forces to be the relevant degrees of freedom and
then model the dynamics of these variables by modifying the stochastic
equations of these modes in the presence of the periodic forces.

Let us first recollect how the dynamics near equilibrium is described in
the absence of driving. In this case the variables of interest, those that
relax very slowly, are the order parameter and the conserved
quantities, if any. This is a small set of variables which are some
functions of the original degrees of freedom. In principle, the dynamics
of these relevant variables can be obtained from the equations of motion
of the original degrees, by integrating out the unwanted variables. In
practice, these dynamical equations for the relevant degrees are not thus
established as they are not usually derivable due to the large number
of unwanted variables involved. Hence these equations are based on certain
guiding principles\cite{hohenberg}: 
({\it i}) Forgoing a large number of unwanted variables renders a
stochastic evolution to the relevant ones.  
({\it ii}) These stochastic dynamical equations should be such that the
large-time distribution gives the right static thermodynamic 
properties. This large-time distribution is the distribution either
obtained by integrating out the irrelevant variables from the canonical
equilibrium distribution or found on phenomenological grounds, for
e.g., Landau-Ginzburg (LG) theory near critical point. Finally, the
stochastic equations thus established could explain the dynamic properties
of the relevant variables near equilibrium.

There exists many different families of stochastic equations that give the
same static properties for the relevant degrees. One of these is the
following Langevin equation, the time-dependent Landau-Ginzburg (TDLG)
theory that describes the dynamics near equilibrium, 
\begin{equation}
\label{tdlg}
\Gamma ~ \frac {\partial} {\partial t}\varphi(x,t) =
- \left. \frac {\delta {\cal H}(\varphi)} {\delta \varphi(x)}  
\right|_{\varphi(x) \rightarrow \varphi(x,t)}
+ \eta(x,t),
\end{equation}
where 
$\varphi(x,t)$ is a $d+1$ dimensional stochastic field that relaxes
to the order parameter $\varphi(x)$ of the system and 
${\cal H}(\varphi)$ is the LG free-energy functional;  
$\eta (x,t)$ is a Gaussian random field with 
$\langle \eta(x,t)\rangle_{\eta} = 0$ and
$\langle \eta(x,t)\eta(x',t')\rangle_{\eta} = 2\Gamma 
\delta(x-x')\delta(t-t')$. 
If $\varphi_{\eta}(x,t))$ is the solution of the above Langevin equation
then the probability distribution of this solution,
$P(\varphi,t) = \prod_x \langle 
\delta (\varphi(x)- \varphi_{\eta}(x,t))\rangle_{\eta}$, evolves 
according to a Fokker-Planck (FP) equation, 
$\partial_t P = {\cal L} P$, that is obtained from
Eq.(\ref{tdlg}) and the Gaussian distribution of $\eta(x,t)$.
The formal solution of this FP equation is
$P(\varphi,t) = \exp({\cal L}t) P(\varphi,t=0)$, where $P(\varphi,t=0)$
is the initial distribution. The asymptotic distribution,
$P_{\infty}(\varphi,t)= \lim_{t \rightarrow \infty} P(\varphi,t)$,
is the right eigenfunctions of ${\cal L}$ operator with zero eigenvalue.
This asymptotic probability distribution is the LG measure, 
$P_{eq}(\varphi) \sim \exp\bigl(-{\cal H}(\varphi)\bigr)$,
provided 
({\it i}) the eigenvalues of ${\cal L}$ are negative semi-definite, 
({\it ii}) there exists a unique normalizable eigenfunction corresponding
to zero eigenvalue in the connected space of functions to which the
initial distribution belongs to, and
({\it iii}) the initial distribution is not orthogonal to this
eigenfunction.

The presence of driving fields on the system will alter both 
its kinematics and dynamics. Assumptions about kinematics are:
({\it i}) The set of variables that were relevant in the absence of
driving for the dynamics near equilibrium continue to remain relevant upon
introducing driving. ({\it ii}) Some  minimal set of additional variables
might become relevant too. We will consider a velocity field
$\pi(x)$ along with the order parameter field $\varphi(x)$  to be 
relevant. The dynamical variables of the system are then specified by
$\varphi(x,t)$ and its time derivative $\partial_t \varphi(x,t)$.
Assumptions about the dynamics are:   
({\it i}) The Langevin equation, that describes dynamics near equilibrium
in the absence of driving, gets modified by adding terms related to the
periodically driven fields and by adding a second-order time-derivate term 
$\partial^2_t\varphi(x,t)$ whose effect is significant when the driving
frequency is high. 
({\it ii}) The periodic driving does not change the properties of the
noise field for any frequency of the driving fields.

The modified TDLG theory that describes the system when subjected to
driving is then given by the following stochastic equation
\begin{equation}
\label{mtdlg}
m\frac {\partial^2 } {\partial t^2} \varphi(x,t) = 
-\Gamma ~ \frac {\partial} {\partial t}\varphi(x,t) 
- \frac {\delta {\cal H}(\varphi)} {\delta \varphi(x,t)}  
+ {\cal F}\bigl(\varphi(x,t),t\bigr) + \eta (x,t),
\end{equation}
where ${\cal F}(\varphi(x),t)$ is the driving field that is periodic
in time with period $T=2\pi/\Omega$. The general form of this field is
\begin{equation}
\label{Fform}
{\cal F}\bigl(\varphi(x),t\bigr) = 
\sum_{n=1}^{\infty} 
\biggl[ F_n\big(\varphi(x)\bigr) \cos(n \Omega t) 
+ G_n\bigl(\varphi(x)\bigr) \sin(n \Omega t) \biggr],
\end{equation}
where $F_n$ and $G_n$ are arbitrary local functionals of $\varphi(x)$.
This term could be thought of as a result of making the coupling constants
in ${\cal H}(\varphi)$ time dependent and periodic. We will take
$m=1$ so that the expressions look simple and to regain the $m$
dependence back replace $t \rightarrow t/\sqrt{m}$,
$\Gamma \rightarrow \Gamma/\sqrt{m}$ and 
$\Omega \rightarrow \sqrt{m}\Omega$. The distribution of the noise field
is Gaussian as specified earlier. 

The above dynamical equation for a $0+1$ dimensional stochastic field
describes periodically driven Brownian particle. These Brownian particles
exhibit a variety of interesting asymptotic behavior depending on the
driving forces\cite{barma}: The particles when driven could
congregate around more than one point even though when in equilibrium they
would have congregated around a single point. Particles with different
masses respond to driving differently and thus when they are mixed and
driven would cluster around different points.

The phase space probability distribution is defined as    
$ P(\varphi,\pi,t) =  \left\langle \prod_x
\delta(\varphi(x)-\varphi_{\eta}(x,t))
\delta(\pi(x)-\partial_t \varphi_{\eta}(x,t))
\right\rangle_{\eta} $, 
where $\varphi_{\eta}(x,t)$ is the solution of Eq.(\ref{mtdlg})
at time $t$ for a particular history of $\{\eta(x,t)\}$ over a time $t$
and $<\cdots>_{\eta}$ is the average over the noise distribution. 
This distribution satisfies the normalization condition 
$\int {\cal D}[\varphi,\pi] P(\varphi,\pi,t) = 1$.
The time evolution of $ P(\varphi,\pi,t)$ is
described by the FP equation
\begin{equation}
\label{FPeqn}
\frac{\partial }{\partial t} P(\varphi,\pi,t)
= {\cal L}(t) P(\varphi,\pi,t),
\end{equation}
where
\begin{eqnarray}
\label{FPop}
{\cal L}(t)
&=& \int_x \left[
-\frac{\delta}{\delta \varphi}\pi
-\frac{\delta}{\delta \pi} 
\left(-\Gamma \pi 
- \frac {\delta {\cal H}} {\delta \varphi}  
+ {\cal F}(\varphi,t)
\right)
+ \Gamma\frac{\delta^2 }{\delta \pi^2}
\right] \nonumber \\
&=& 
{\cal L}_{FP} -\int_x \frac{\delta}{\delta \pi} {\cal F}(\varphi,t)~.
\end{eqnarray}
The FP operator ${\cal L}_{FP}$ is defined as the time-independent
part of the ${\cal L}(t)$ operator, both of which are assumed to be
suitably regularized. To keep the notation compact, $\varphi =\varphi(x)$
and $\pi=\pi(x)$ are used in the above equation. The solution of this
equation is not known in general and our aim is to find the asymptotic
solution for some range of parameters of the system and driving fields.

The asymptotic distribution of the above FP equation is a periodic
function with period $T$, if the real part of the eigenvalues of
$\partial_t - {\cal L}(t)$ are positive definite. In brief, the argument
for the periodicity goes as follows. The solution of the FP equation is
the right eigenfunction of the operator $\partial_t - {\cal L}(t)$
corresponding to the zero eigenvalue. If ${\cal L}(t)= {\cal L}(t+T)$ then 
$\partial_t - {\cal L}(t)$ commutes with the discrete time-translation
operator $\exp{(T\partial_t)}$. The solution can then be expanded in terms
of the common right eigenfunctions of these two operators. Let the
eigenfunctions of $\partial_t - {\cal L}(t)$ and $\exp{(T\partial_t)}$
with eigenvalues $0$ and $\exp{(\mu T)}$, respectively, be the
Floquet-type functions $\exp{(\mu t)} p_{\mu}(t)$, 
where $p_{\mu}(t)$ is a periodic function with period $T$.
Substituting these eigenfunctions in FP equation gives 
$[\partial_t - {\cal L}(t)]p_{\mu}(t)= -\mu p_{\mu}(t)$. Hence, if the
real part of the eigenvalues of $\partial_t - {\cal L}(t)$ are positive
definite then in the limit $t \rightarrow \infty$ the only eigenfunction 
that survives is $p_{0}(t)$, thus making the asymptotic distribution
periodic.

When the coupling constants $\{ \tilde{g} \}$ of the driving field
${\cal F}(\varphi,t)$ are small compared to the coupling constants 
$\{ g \}$ of ${\cal H}(\varphi)$ then the FP equation can 
be solved perturbative in $\{ \tilde{g} \}$ provided we know the right and
left eigenfunctions of the FP operator ${\cal L}_{FP}$. These
eigenfunctions are generically not known though the eigenfunctions of the
FP operator which includes only the free part of ${\cal H}(\varphi)$ are
obtainable. Hence, the eigenfunctions of ${\cal L}_{FP}$  can be
determined perturbatively in $\{ g \}$ and in turn the solution of the FP
equation in the double series expansion in 
$\{ {g} \}$ and $\{ \tilde{g} \}$.

We will now further to find the asymptotic behavior of the FP equation
when $\{ \tilde{g} \}$ are not necessarily small compared to $\{ {g} \}$. 

\section{Asymptotic distribution of the FP equation}
\label{asymp}
In this section, we will first transform the FP equation to enable us 
to have a non-perturbative solution in $\{ \tilde{g} \}$ but perturbative
in a parameter that involves both $\{ \tilde{g} \}$ and 
$\Omega^2 + \Gamma^2$. We then explicitly evaluate the asymptotic
distribution to first order in this parameter and more specifically the
time-averaged asymptotic correlation functions. We find that these
correlation functions can be expressed as equilibrium correlation
functions with an effective LG energy functional.

\subsection{Formal solution}
\label{formal}
We will make a change of variables and transform the FP equation such that
the time-dependent part of the transformed FP operator is small
though the time-dependent part of the original FP equation is not. 
The variables $\{\varphi, \pi, t \}$ are changed to $\{\Phi, \Pi, \tau \}$
by the following nonlinear transformation
\begin{eqnarray}
\varphi(x)&=& \Phi(x) + \xi (x;\Phi,\tau), \nonumber \\
\pi(x)&=& \Pi(x) + \partial_{\tau}\xi (x;\Phi,\tau), \nonumber \\
t&=&\tau,
\end{eqnarray}
where the explicit form of $\xi(x;\Phi,\tau)$ will be specified later. 
Under this transformation the probability distribution is made to
behave like a scalar:
$P(\varphi, \pi, t) \rightarrow
\widetilde{P}(\Phi, \Pi, \tau)= P(\varphi, \pi, t)$.
The functional derivatives will then transform as 
\begin{eqnarray}
\frac{\delta}{\delta \pi(x)} 
&=& \frac{\delta}{\delta \Pi(x)}, \nonumber \\
\frac{\delta}{\delta \varphi(x)} 
&=& \int_y D(x,y) 
\left( \frac{\delta}{\delta \Phi(y)} 
- \int_z  \partial_{\tau} M(y,z)\frac{\delta}{\delta \Pi(z)} 
\right),
\end{eqnarray}
and the time derivative as
\begin{eqnarray}
\frac{\partial}{\partial t}=
\frac{\partial}{\partial \tau}
-\int_{x} \partial^2_{\tau} \xi (x;\Phi,\tau) 
\frac{\delta}{\delta \Pi(x)}
-
\int_{x} \partial_{\tau} \xi (x;\Phi,\tau) 
\frac{\delta}{\delta \varphi(x)} ,
\end{eqnarray}
where
\begin{eqnarray}
M(x,y)=\frac{\delta} {\delta\Phi(x)}\xi(y;\Phi,\tau), ~
\int_y D(x,y) \frac{\delta\varphi(z)}{\delta\Phi(y)}= \delta(x-z).
\end{eqnarray}
Substituting the above new variables in Eq.(\ref{FPeqn}) we obtain the
following transformed FP equation 
\begin{eqnarray}
\label{TFPeqn}
\frac{\partial \widetilde{P} } {\partial t} 
&=& \int_x \left[
-\frac{\delta}{\delta \Phi} \Pi
-\frac{\delta}{\delta \Pi} 
\bigl(-\Gamma \Pi 
+ f_{\cal H} (\Phi + \xi)
+ \Delta{\cal F}(\xi, \tau)
\bigr)
+ \Gamma\frac{\delta^2 }{\delta \Pi^2}
\right] \widetilde{P}
\nonumber \\
&+&
\int_{x,y}\Pi(x) \left[
\biggl\{ \delta(x-y)- D(x,y)\biggr\} 
\frac{\delta}{\delta \Phi(y)} 
+ D(x,y) \int_{z} \partial_{\tau}M(y,z)
\frac{\delta}{\delta \Pi(z)} \right] \widetilde{P} \nonumber \\
&+& 
\int_x
\left[\biggl( \partial^2_{\tau}\xi + \Gamma \partial_{\tau}\xi
- {\cal F}(\Phi, \tau)\biggr)  \frac{\delta}{\delta \Pi(x)}\right] 
\widetilde{P},
\end{eqnarray}
where 
$f_{\cal H} (\varphi(x)) = - {\delta {\cal H}}/ {\delta \varphi(x)}$
and
$\Delta{\cal F}(\xi, \tau) =
{\cal F}(\Phi + \xi, \tau) - {\cal F}(\Phi, \tau)$.
We now choose $\xi (x;\Phi,\tau)$ to be the solution of the following
equation 
\begin{equation}
\label{xieqn}
 \partial^2_{\tau}\xi + \Gamma \partial_{\tau}\xi
- {\cal F}(\Phi, \tau) =0,
\end{equation}
whose explicit form is
\begin{eqnarray}
\label{xisoln}
\xi(x;\Phi,\tau)
= \sum_{n=1}^{\infty}
\frac{-1}{n^2\Omega^2 + \Gamma^2}
\left[ \left( F_n\bigl(\Phi(x)\bigr)+\frac{\Gamma}{ n \Omega} 
G_n\bigl(\Phi(x)\bigr) \right)\cos(n\Omega \tau)
+ 
\left( G_n\bigl(\Phi(x)\bigr)
-\frac{\Gamma}{n \Omega}F_n\bigl(\Phi(x)\bigr)
\right) \sin(n\Omega\tau)
\right].
\end{eqnarray}
The reason for this specific choice for $\xi$ is that not only does it
make the last term on the right hand side of Eq.(\ref{TFPeqn}) zero but
also makes the (time)$\tau$-dependent part of the modified FP operator to
be of $O(\xi)$. Thus if $\xi$ is small then this modified equation will
yield a perturbative solution. Upon substituting $\xi$ from
Eq.(\ref{xisoln}) in Eq.(\ref{TFPeqn}) we finally get 
\begin{equation}
\label{fTFPeqn}
\frac{\partial}{\partial \tau} \widetilde{P} (\Phi, \Pi, \tau)
= \bigl[{\cal L} + \Delta{\cal L}\bigr] 
\widetilde{P} (\Phi, \Pi, \tau),
\end{equation}
where ${\cal L}$ is the following static FP operator
\begin{eqnarray}
\label{defL}
{\cal L} =
\int_x \left[
-\frac{\delta}{\delta \Phi} \Pi
-\frac{\delta}{\delta \Pi} 
 \left(-\Gamma \Pi 
+ \overline{f_{\cal H} (\Phi + \xi)}
+ \overline {\Delta{\cal F}(\xi, \tau)}
\right)
+ \Gamma\frac{\delta^2 }{\delta \Pi^2}
\right], 
\end{eqnarray}
and $\Delta {\cal L}$ is the periodic (time)$\tau$-dependent operator
\begin{eqnarray}
\label{defDL}
\Delta {\cal L} &=&
\int_x \left[
-\frac{\delta}{\delta \Pi} 
\left( 
f_{\cal H} (\Phi + \xi) - \overline{f_{\cal H} (\Phi + \xi)}
+\Delta{\cal F}(\xi, \tau) - \overline {\Delta{\cal F}(\xi, \tau)}
\right)
\right]  \nonumber \\
&+&
\int_{x,y}\Pi(x) \left[
\biggl\{ \delta(x-y)- D(x,y)\biggr\} \frac{\delta}{\delta \Phi(y)} 
+ D(x,y) \int_{z} \partial_{\tau}M(y,z)
\frac{\delta}{\delta \Pi(z)} \right],
\end{eqnarray}
where the {\it bar} over the terms indicate an average over a time period.

The perturbative asymptotic solution can be formally written as
\begin{eqnarray}
\label{fsol}
\widetilde{P}_{\infty} (\Phi, \Pi, \tau)
= Q (\Phi, \Pi)
+ \frac{1}{\partial_{\tau} -{\cal L}} 
\Delta{\cal L} \widetilde{P}_{\infty} (\Phi, \Pi, \tau),
\end{eqnarray}
where $Q (\Phi, \Pi)$ is the right eigenfunction of ${\cal L}$ with zero
eigenvalue. If ${\cal L}$ has a unique eigenfunction corresponding to
zero eigenvalue, and the real part of the non-zero eigenvalues do not
vanish, and $\widetilde{P}_{\infty}$ is periodic in time,
then it follows from Eq.(\ref{fTFPeqn}) that
$\Delta{\cal L}\widetilde{P}_{\infty}$ has no overlap with the
eigenfunction of ${\partial_{\tau} -{\cal L}}$ corresponding to zero
eigenvalue. Thus, though ${\partial_{\tau} -{\cal L}}$ is not invertible,
its inverse action on the space orthogonal to its eigenfunction with zero
eigenvalue is well defined.

\subsection{Effective theory}
\label{eftheory}
In this subsection, we will obtain the effective energy functional 
to $O(\xi)$ by averaging the asymptotic correlation functions
over a time-period. The observables will be related to the time-averaged
correlation functions if the time of observation is comparable to the
time period of the driving fields. In other words, we are assuming that
the driving fields oscillate rapidly compared to the time scale of
the measurement.

The equal-time correlation functions of the stochastic field transform
under the change of variables as follows
\begin{eqnarray}
\label{corr}
\langle \varphi(x_1,t) \varphi(x_2,t) \cdots \rangle_{\eta}
&=& \int {\cal D}[\pi,\varphi]
\varphi(x_1) \varphi(x_2) \cdots P(\varphi,\pi,t) \nonumber \\
&=& \int {\cal D}[\Pi,\Phi] J[\Phi]
\biggl\{\Phi(x_1) + \xi (x_1;\Phi,t)\biggr\}
\biggl\{\Phi(x_2) + \xi (x_2;\Phi,t)\biggr\}\cdots 
\widetilde{P}(\Phi,\Pi,t),
\end{eqnarray}
where the integration measure
${\cal D}[\pi,\varphi]= \prod_x [d\pi(x)d\varphi(x)]$ 
and the Jacobian of the transformation
$J[\Phi] =\prod_x [1+ {\partial \xi\bigl(x,\Phi(x),t\bigr)}
/ {\partial \Phi(x)}]$. The first-order asymptotic distribution, obtained
from Eq.(\ref{fsol}), is  
$
\widetilde{P}_{\infty}= Q^{(1)}
+ (\partial_{\tau} -{\cal L})^{-1} 
\Delta{\cal L} Q^{(0)}
$,
where $Q^{(0)}$ and $Q^{(1)}$ are the solutions of the equation 
${\cal L}Q=0$ to $O(1)$ and $O(\xi)$, respectively. Substituting 
$\widetilde{P}_{\infty}$ in Eq.(\ref{corr}) and averaging over a period
gives
\begin{eqnarray}
\label{meancorr}
\int {\cal D}[\pi,\varphi]
\varphi(x_1) \varphi(x_2) \cdots \overline{P_{\infty}[\varphi,\pi,t]} 
&=& \int {\cal D}[\Pi,\Phi] 
\Phi(x_1)\Phi(x_2)\cdots Q^{(1)}(\Phi,\Pi)
+ O(\xi^2),
\end{eqnarray}
since the time-average of both $\xi$ and
$[(\partial_{\tau} -{\cal L})^{-1} \Delta{\cal L} Q^{(0)}]$ are zero.
From Eq.(\ref{defL}), in which $\overline{f_{\cal H} (\Phi + \xi)}$
and $\overline{\Delta{\cal F}(\xi, \tau)}$ are expanded to $O(\xi)$, 
we get
\begin{eqnarray}
\label{Q1}
Q^{(1)}(\Phi,\Pi)= \frac{1}{Z^{(1)}}
\exp\biggl( -\frac{1}{2}\int_x \Pi^2(x) 
- {\cal H}(\Phi)- \Delta{\cal H}(\Phi)\biggr),
\end{eqnarray}
where $Z^{(1)}$ is the normalization constant and $\Delta{\cal H}(\Phi)$
satisfies the condition
\begin{eqnarray}
\label{DelH}
\frac{\delta}{\delta \Phi(x)}\Delta{\cal H}(\Phi)
= - \overline{\xi(x;\Phi,t) 
\frac{\partial }{\partial \Phi(x)} {\cal F}\bigl(\Phi(x),t\bigr) }.
\end{eqnarray}
Substituting $\xi$ and ${\cal F}$ from Eqs. (\ref{xisoln}) and
(\ref{Fform}) in the above equation we get 
$\Delta{\cal H}=\Delta{\cal H}_1+\Delta{\cal H}_2$, where 
\begin{eqnarray}
\label{DH}
\Delta{\cal H}_1(\Phi) &=& \frac{1}{4}
\sum_{n=1}^{\infty}
\frac{1}{n^2\Omega^2 + \Gamma^2}
\int_x
\biggl( F_n^2\bigl(\Phi(x)\bigr) + G_n^2\bigl(\Phi(x)\bigr)\biggr),
\nonumber \\
\Delta{\cal H}_2(\Phi)&=&  
\sum_{n=1}^{\infty}
\frac{\Gamma}{(n^2\Omega^2 + \Gamma^2)(2n \Omega)}
\int_x {\cal W}\bigl[F_n\bigl(\Phi(x)\bigr),G_n\bigl(\Phi(x)\bigr)\bigr].
\end{eqnarray}
${\cal W}\bigl[F_n\bigl(\Phi(x)\bigr),G_n\bigl(\Phi(x)\bigr)\bigr]$ 
is defined by the indefinite integral 
${\cal W}[F(\alpha), G(\alpha)]=\int d\alpha
[G(\alpha)F'(\alpha)-F(\alpha)G'(\alpha)]$. After integrating out $\Pi$ in
Eq.(\ref{meancorr}), we finally get
\begin{eqnarray}
\label{effcorr}
\overline{
\lim_{t \rightarrow \infty}
\langle \varphi(x_1,t) \varphi(x_2,t) \cdots \rangle_{\eta} }
= \frac{1}{Z} \int {\cal D}[\varphi(x)]
\varphi(x_1) \varphi(x_2) \cdots
e^{-{\cal H}_{eff}(\varphi)}, 
\end{eqnarray}
where $Z=\int{\cal D}\varphi \exp \bigl(-{\cal H}_{eff}\bigr)$
and ${\cal H}_{eff}(\varphi)={\cal H}(\varphi)+\Delta{\cal H}(\varphi)$.
The effective energy functional for a $N$-component
order parameter field $\varphi \equiv \{ \varphi_a \}$ is a
straight-forward generalization of Eqs. (\ref{DelH}) and (\ref{DH}).
In this case $\Delta{\cal H}$, which satisfies
Eq.(\ref{DelH}) for each component, exists only if the components
of the driving field $F^a_n$ and $G^a_n$ satisfy the conditions, 
$\partial F^a_n/ \partial \varphi_b = \partial F^b_n/ \partial \varphi_a$
and
$\partial G^a_n/ \partial \varphi_b = \partial G^b_n/ \partial \varphi_a$.

Note that we have neglected $O(\xi^2)$ terms in the expansion of the
${\cal L}$ and $\Delta {\cal L}$ operators because $\xi$ is assumed
to be small. The condition for the smallness of $\xi$ and the criteria for
the validity of $O(\xi)$ approximation are both obtained by comparing the
neglected terms with those that are retained in these operators. For
instance, if ${\cal F}= 2 \tilde{g} \varphi^2(x) cos(\Omega t)$ and the
coefficient of $\varphi^4$ term in ${\cal H}(\varphi)$ is $\lambda$ then
$\xi$ is small if $\lambda \gg \tilde{g}^2/(\Omega^2+\Gamma^2)$. A
self-consistent criteria can also be obtained by comparing the expectation
values of the neglected and the retained terms using ${\cal H}_{eff}$. In
some cases the higher powers of $\xi$ contribute only irrelevant terms to
${\cal H}_{eff}$ and hence can be neglected while determining universal
properties, independent of $\xi$ being small. 
For instance, if ${\cal F}= \tilde{g} \varphi^4(x) cos(\Omega t)$ and
${\cal H}$ is a $\varphi^4$ theory in four dimensions, then the $O(\xi)$
term in ${\cal H}_{eff}$ itself is an irrelevant $\varphi^8$ field and 
the higher-order terms contain fields that are more irrelevant.

The effective energy functional ${\cal H}_{eff}$ can be interpreted as
follows. The asymptotic distribution contains both statistical
fluctuations $\Phi$ and the dynamical fluctuations $\xi(\Phi,t)$ that are
periodic in time. In Eq.(\ref{DelH}), if we substitute ${\cal F}$ in terms
of $\xi$ by using Eq.(\ref{xieqn}) and then integrate by parts, we get   
\begin{eqnarray}
\label{meaning}
\frac{\delta}{\delta \Phi}\Delta{\cal H}(\Phi)
= \frac{\partial }{\partial \Phi}
\frac{1}{2}\overline{ \bigl( \partial_t\xi \bigr)^2 }
+ \Gamma \overline { \partial_t\xi 
\frac{\partial }{\partial \Phi} \xi }.
\end{eqnarray}
Thus, by averaging over the time period we have eliminated the dynamical
fluctuations and provided an effective description to the system in
terms of the statistical fluctuations that are governed by a modified
energy functional ${\cal H}_{eff}={\cal H}+\Delta{\cal H}$.

\section{Illustrative examples}
\label{examples}
We have seen that the large-time behavior of periodically driven
stochastic systems, averaged over a period, can be described by an
effective LG functional of equilibrium systems. We can now make use of
the knowledge about these equilibrium systems to induce new phases and
phase transitions in various systems by subjecting them to time-dependent
periodic fields. We illustrate some of the effects due to these driving
fields with a few examples.

\subsubsection{Varying the critical point}

The driving fields transform the energy functional ${\cal H}$ into 
a new energy functional ${\cal H}_{eff}$. This amounts to changing the
coupling constants, $\{ g\} \rightarrow \{g_e \}$,
which in turn induces  a change in the critical point. We will illustrate
this with an example. The LG functional that describes the Ising model
near the critical temperature is 
\begin{equation}
\label{eg1H}
{\cal H}[\varphi]= \int_x \bigl(\partial\varphi(x)\bigr)^2 + 
A(T-T_c)\varphi^2(x) + \lambda \varphi^4(x). 
\end{equation}
The mean-field theory suggests a second order transition at $T_c$ from a 
$Z_2$-symmetry-broken phase to the symmetry-unbroken phase and a
dependence of magnetization (order parameter) on temperature below $T_c$
as $\langle \varphi \rangle^2=A(T_c-T)/2\lambda$. If we now drive this
system, say, by oscillating $\varphi^2$ term, which is same as adding
a force term 
\begin{equation}
{\cal F}\bigl(\varphi(x)\bigr) =2\tilde{a}\varphi(x)\cos(\Omega t),
\end{equation}
then this system will get described by the following effective energy
functional that is obtained using Eq.(\ref{DH}) 
\begin{equation}
{\cal H}_{eff}[\varphi]= \int_x \bigl(\partial\varphi(x)\bigr)^2 + 
a_e \varphi^2(x) + \lambda \varphi^4(x), 
\end{equation}
where $a_e = A(T-T_c) +\tilde{a}^2/(\Omega^2+\Gamma^2)$.
Since this effective energy functional differs from the original
functional only in the coefficient of the $\varphi^2$ term one can read
off the critical temperature $\theta_c$ and the behavior of the
time-averaged magnetization $\overline{\langle \varphi \rangle}$ 
below $\theta_c$ of the driven system. We get
\begin{eqnarray}
\theta_c &=& T_c - \frac {\tilde{a}^2} {A(\Omega^2+\Gamma^2)},
\nonumber \\
\overline{\langle \varphi \rangle}^2 &=& \frac {A} {2\lambda}
(\theta_c-T).
\end{eqnarray}
Hence, the driving field acting on this system tends to destroy the
symmetric phase as it reduces both the critical temperature and the
magnetization at a given temperature.

\subsubsection{Changing the nature of transition}

The nature of phase transition can also be changed by applying driving
fields. Consider the LG functional as given in the previous example,
Eq.(\ref{eg1H}), but with a different driving force 
\begin{equation}
{\cal F}\bigl(\varphi(x)\bigr)= 2\bigl({\tilde a}\varphi(x)- 
{\tilde b} \varphi^3(x) \bigr) \cos(\Omega t),
\end{equation}
and also assume the dimension of space $d > 3$ where
$\varphi^6$ term is irrelevant. The effective energy functional will
then become
\begin{equation}
{\cal H}_{eff}[\varphi]= \int_x \bigl(\partial\varphi(x)\bigr)^2 + 
a_e \varphi^2(x) + \lambda_e \varphi^4(x) + b_e \varphi^6(x),  
\end{equation}
with the coupling constants
$ a_e=A(T-T_c) + {\tilde a}^2/(\Omega^2+\Gamma^2)$,
$\lambda_e=\lambda - 2{\tilde a}{\tilde b}/(\Omega^2+\Gamma^2)$, and
$b_e={\tilde b}^2/(\Omega^2+\Gamma^2)$.

Now the mean-field prediction is as follows. For $\lambda_e > 0$ two
different phases exist; a spontaneously broken phase when $a_e$ is
negative and an unbroken phase when $a_e$ is positive. These
phases are separated by a line of second order transition which ends in a 
tricritical point when both $\lambda_e$ and $a_e$ become zero. For
$\lambda_e < 0$ there are three possibilities depending on the value of
$a_e$: ({\it i}) If $a_e< \lambda_e^2/4b_e $ then it is in a symmetry
broken phase. ({\it ii}) If $\lambda_e^2/4b_e <a_e< \lambda_e^2/3b_e $
then it is a metastable phase and will have a broken symmetry if one
enters this region crossing the line $\lambda_e^2=4a_eb_e$ and will
have an unbroken symmetry if one enters by crossing the line 
$\lambda_e^2=3a_eb_e$. ({\it iii}) If $a_e> \lambda_e^2/3b_e $ then it is
in the symmetric phase. The line $\lambda_e^2=4a_eb_e$ is a line of
first-order transition which ends in the tricritical point. 

Suppose the system initially is in the symmetric phase with both
$a_e$ and $\lambda_e$ positive. Now switch on the periodic force such
that the product ${\tilde a}{\tilde b}$ is positive and then gradually
reduce the driving frequency $\Omega$. This will reduce both $a_e$ and
$\lambda_e$ and takes the system across the line of first-order transition
into the symmetry broken phase. More generally, by tuning at most two of
the three parameters $\{{\tilde a},{\tilde b},\Omega\}$ and the
temperature $T$ we can scan the entire $a_e-\lambda_e$ plane. Hence, by
applying the driving fields we can engineer the behavior of the system.  

\subsubsection{Inducing new fixed points}

Driving fields can enlarge the coupling constant space of the system
by introducing either relevant or irrelevant fields. Though the irrelevant
fields usually do not change the large-distance properties of the system
but in some cases, like in the previous example, the coupling constant of
the relevant fields can get shifted to a region where irrelevant fields
become important. Let us now examine an example where the driving fields
introduce a relevant field and drastically alter the large-distance
properties. This is because the system flows under scaling to a new stable
fixed point in the enlarged coupling constant space. The system under
study is described by the $O(N)$ model  
\begin{equation}
{\cal H}= \int_x \bigl(\partial \varphi(x)\bigr)^2 + t \varphi^2(x) + 
u \bigl(\varphi^2(x)\bigr)^2,
\end{equation}
where $\varphi \equiv \{\varphi_1, \cdots, \varphi_N\}$ is a $N$-component
vector field. This model has two fixed points: one is at $u=0$ (Gaussian
fixed point) and the other is at finite $u$ (Heisenberg fixed point). For
dimension $d < 4$ Heisenberg fixed point is stable while Gaussian is not.
Now drive the system by the fields 
\begin{equation}
{\cal F}_a\bigl(\varphi(x)\bigr) = 2 \varphi_a^2(x) \cos(\Omega t), 
\end{equation}
$a= 1,\cdots, N$. The resultant model is the $O(N)$ model with a cubic
symmetry breaking term\cite{aharony}: 
\begin{equation}
{\cal H}_{eff} = {\cal H} + v \int_x \sum_a \varphi_a^4(x), 
\end{equation}
where 
$v= (\Omega^2 + \Gamma^2)^{-1}$. In the $u-v$ plane this model has two
more fixed points, Ising fixed point at $(u=0, v\neq 0)$ and the 
cubic fixed point at $(u\neq 0, v\neq 0)$, apart from the  Gaussian and
Heisenberg fixed points at $(u=0, v=0)$ and $(u\neq 0, v=0)$,
respectively. The stability of the fixed points is as
follows: The Gaussian point is unstable in the $v$-direction too. Ising
point is stable in the $v$-direction but unstable in the $u$-direction.
Heisenberg point is stable in the $v$-direction if $N < 4$ and unstable if
$N > 4$. Cubic point is stable in both directions if Heisenberg point is
unstable and vice-versa. Thus, when $N > 4$ these driving fields change
the large-distance properties of the system from Heisenberg to cubic.
Periodically driven stochastic models with $O(N)$ symmetry, without the
$m\partial^2_t\varphi(x,t)$ in Eq.(\ref{mtdlg}), were first studied in the
context of magnetic hysteresis\cite{rao,dhar}. The phase transitions that
get induced by the driving fields in these models  were recently 
investigated too\cite{kurchan}.   

In summary, we have derived the effective theory for the correlation
functions of a class of periodically driven macroscopic systems. We have
shown with a few examples that this effective theory can be made use to
select the driving fields that can steer the system through a plethora of
phases.

\acknowledgments
It is a pleasure to thank Mustansir Barma and Deepak Dhar for helpful
comments and useful discussions. This work was supported in toto by the
people of India.

\end{document}